\documentclass[12pt]{article}

\def\xyma{\xymatrix@M.7em}
\def\xymas{\xymatrix@M.1em}

\begin{document}
\thispagestyle{empty}

\begin{titlepage}
\begin{center}
\hfill SU-ITP-05/01\\
\hfill DFPD-05/TH/02\\
\hfill {\tt hep-th/0501081}\\

\

\bigskip
\begin{center}
{\large \bf   Dirac Action on M5 and M2 Branes with Bulk Fluxes}

\end{center}

\vskip 1cm

{\bf
Renata Kallosh$^{1}$
and Dmitri Sorokin$^{2}$
}
\\

\vskip 1cm


{\em }

\

\centerline{$^1$ Institute for Theoretical Physics, Department of Physics}
\centerline{Stanford University, Stanford, CA 94305}
\centerline{ {\tt kallosh@stanford.edu} }

\centerline{$^2$ INFN Sezione di Padova ${\&}$ Dipartimento di
Fisica ``Galileo Galilei", } \centerline{Universit\`{a} degli Studi
di Padova, 35131, Padova, Italy}
 \centerline{\it {\tt
dmitri.sorokin@pd.infn.it} }

\end{center}

\

\begin{abstract}
We derive an explicit form of the quadratic-in-fermions  Dirac
action on the M5 brane for an arbitrary on-shell background of 11D
supergravity with non-vanishing fluxes and in presence of a chiral
2-form on M5.  This action may be used to generalize
the conditions for which the non-perturbative superpotential  can be
generated  in M/string theory. We also derive the Dirac action with
bulk fluxes on the M2 brane.
\end{abstract}

\vfill

\

January 2005\\
\end{titlepage}

\section{Introduction}

Our purpose is to derive the quadratic action for fermions on the
M5 brane in a background of 11D supergravity with generic fluxes.
Although  the action of M5  in supergravity background
\cite{Bandos:1997ui} as well as the equations of motion for all
fields on M5 surface are known
\cite{Howe:1997fb,Bandos:1997gm,Sorokin:1999jx}, an explicit
dependence on 4-form fields of 11D supergravity in the Dirac
action on M5 has not been given yet.

The motivation for this work is two-fold. First of all to find an
explicit form of a Dirac operator in presence of generic fluxes for
the fundamental brane objects  may have its own merit, especially if
the answer is reasonably simple and allows to understand
implications for the physics of the branes in such backgrounds. For
the M2 and Dp branes such studies have already been undertaken in
the literature. For instance, the coupling of D3 brane fermions to
type IIB supergravity fluxes was studied in {\cite{Grana:2002tu}}.
 And in \cite{Marolf:2003vf,Marolf:2003ye} Dp brane actions in
generic supergravity backgrounds were derived in the quadratic
approximation for fermions (in the Green--Schwarz form without gauge
fixing worldvolume reparametrization and kappa-symmetry) from a
corresponding M2 brane action \cite{Grisaru:2000ij} using T--duality
 \footnote{Other aspects of Dirac operators on branes have been
considered e.g. in \cite{Claus:1997cq}-\cite{Gomis:2004pw}.}. In
this paper we present the results for the M5 brane and also for the
M2 brane both in the Green--Schwarz and purely worldvolume form.
This will allow one to investigate, in particular, various matrix
models in backgrounds which are not necessarily of an
$AdS\times S$-type but more general ones.

Secondly, significant part of the motivation for this work was the recent
interest in non-perturbative corrections to M/string theory. It has
been argued by Witten in \cite{Witten:1996bn} a while ago that  the
$SO(2)$ symmetry, which is a subgroup of the structure group $SO(5)$
may be the exact symmetry of the M5 brane action, no  background
fluxes were considered at that time. This was a basis for
establishing a very powerful theorem about the conditions when
non-perturbative superpotentials can be generated in M/string
theory.

The analysis is based on algebraic geometry and leads to the statement that
the  compactification four-fold
must admit divisors of arithmetic
genus one, $\chi_D \equiv \sum (-1)^{n} h_n=1$. This indicates that
 in type IIB compactifications
there can be {\it non-vanishing corrections to the superpotential} coming from Euclidean
D3 branes.
In the presence of such instantons, there is a correction to the
superpotential which at large volume yields a  term
$
W_{\rm inst} = T(z_i) \exp(2\pi i \rho)
$
where $T(z_i)$ is a one-loop determinant,
and the leading exponential dependence comes from the action of a
Euclidean D3 brane wrapping a four-cycle in the compactified manifold.
The non-perturbative corrections to the superpotential
are absent, according to \cite{Witten:1996bn}, when $\chi_D \neq 1$ (in absence of fluxes).

The presence of such non-perturbative corrections to the superpotential plays a crucial role in the
stabilization of the
volume modulus, as was shown in the simplest KKLT model with one K\"ahler modulus \cite{Kachru:2003aw} and in a
general class of models with many K\"ahler moduli in \cite{Denef:2004dm}. In particular, in all
models of Denef, Douglas and Florea in \cite{Denef:2004dm} the choice of compactification manifolds
was always satisfying the
restriction that $\chi_D=1$ and a significant effort was made to find them.

An analysis of some of these issues was performed in
\cite{Robbins:2004hx} where the role of the 4-flux in the generation
of instanton corrections has been discussed. In particular it was shown that the total flux through the divisor must vanish.

It was suggested  by Gorlich, Kachru, Tripathy and Trivedi
in \cite{Gorlich:2004qm} and argued for particular examples of
compactification that in the presence of fixed fluxes the $U(1)$
symmetry of the fermionic action of the M5 brane might be broken.
This in turn leads to a  possibility of generating
non-perturbative superpotentials in models with divisors on a
four--fold of an arithmetic genus $\chi_D\geq 1$.

 There are  two possible points of view on  $U(1)$ symmetry in
presence of fluxes. These  correspond to two  well known  aspects
 of symmetries of the background functional in field theory
\cite{DeWitt:1967ub}. One is related to the so-called ``quantum
gauge transformations'' and the second one is related to the
``background gauge transformations'' accompanied by the
corresponding transformations of the quantum fields (variables of
integration). In our case  when the $U(1)$ symmetry acts as
``quantum gauge transformations''
 only the fermions on the M5 brane are transformed but background
fluxes are kept fixed. In presence of fixed background fluxes the
corresponding $U(1)$ symmetry on the M5 brane may be broken.  On
the other hand, the
$U(1)$ symmetry acting as the ``background gauge transformations''  in our
case means that  the fluxes, if they are vectors or tensors in
the structure group, transform  simultaneously with the action of
$U(1)$ on fermions. In this  sense  the classical Dirac action on the
brane has an unbroken $U(1)$ symmetry in presence of the background
fluxes.

In the setting used in \cite{Witten:1996bn} we find it useful to
consider  fluxes transforming together with fermions under the
$U(1)$--symmetry so that the classical invariance of  the M5 brane action
takes place. This symmetry of the classical theory  may  be broken
by one-loop anomalies. However,  with account of the inflow from the
bulk these anomalies are expected to be canceled
\cite{Witten:1996bn} since the $U(1)$ is just a part of
diffeomorphisms and the theory is expected to be exactly invariant.

This all development is suggestive that the exact dependence of
the Dirac operator on M5 with fluxes may help to understand the
situation in a completely general setting since with account of
bulk fluxes the algebraic properties of the M5 brane Dirac
operator will change. To find the corresponding changes we need to
find an explicit contribution of bulk fluxes to the Dirac operator
on the M5 brane. We may proceed in two ways.

1. First we may look for the fermion action in a target space
covariant Green-Schwarz form where the corresponding
anti-commuting fields are worldvolume scalars and target space
spinors of $Spin (1,10)$. For the reader who just wants to see the
effect of the flux $F_{\underline a \underline b \underline c
\underline d}$ on M5 and M2 fermions, here is the simplified form
(without tensor field couplings on the M5 brane) of the Dirac
equation for the 16-component `kappa-projected' spinor in the
background with fluxes.   Note that in the approximation without
tensor fields Dirac equations have the same form for
 M5 and M2
\begin{equation}\label{Diracflux1}
 \Gamma^{a}(\nabla_{a} +   T_{a}{}^{\underline
a \underline b \underline c \underline d}\,F_{\underline a
\underline b \underline c \underline d}) \theta_- =0\,.
\end{equation}
Here $T_{a}{}^{\underline a \underline b \underline c \underline
d}$ stands for a product of $\gamma$-matrices (the detailed
notation is introduced below).

2. We may look for the action for the world-surface fermions
transforming in $Spin(1.5)\times Spin (5)$. This form is more
useful for the future studies of the instanton effects from the
Euclidean M5 brane wrapped on a six--cycle of a Calabi--Yau
4--fold \cite{Witten:1996bn}.  We will need eventually to perform
an analytic continuation to Euclidean space with spinors in
$Spin(6)\times Spin (5)$. Here again is the simplified form of the
Dirac equation for the chiral spinor on the M5 brane
\begin{equation}\label{Diracflux2}
\left[{\tilde \gamma}^{a} \,\nabla_a
+{1\over{24}}\,\left(\gamma^{ijk}\,
\tilde\gamma^{a}\,F_{aijk}-\gamma^i\,
\tilde\gamma^{abc}\,F_{abci}\right)\,\theta\right]^{\alpha}_q
=0\,,
\end{equation}
see notation and details in the paper. An analogous equation is also
given for the M2 brane below.

The derivation of the explicit form of the quadratic--in--fermions
Dirac action on the M5 brane for an arbitrary on-shell background
of 11D supergravity with non-vanishing fluxes and tensor fields on
the M5 brane and on the M2 brane and of the corresponding
equations of motion is explained below.

 Since we are looking for a quadratic
part of the action in a non-trivial background we may simply look
at fermionic equations of motion which in a general form can be
obtained either using superembedding techniques \cite{Howe:1997fb}
or directly from the M5--brane action
\cite{Bandos:1997ui,Bandos:1997gm}. We will follow notation of the review paper
\cite{Sorokin:1999jx}.

In Sec. 2 we derive the Green-Schwarz type fermionic equations of
the M5 brane in a flux background  for space-time spinors being
world-volume scalars. In Sec. 3 we derive Dirac equations for the
world-volume spinors in a flux background. In both cases
$\kappa$-symmetry is effectively gauge-fixed, so that the relevant
fermion is 16-component. In Sec. 4 we present the Dirac action on M5
in a flux background with  examples of how  $U(1)$ symmetry acts
on the fluxes. Dirac equations and the corresponding Lagrangian
with bulk fluxes on the M2 brane are given in Sec. 5. In conclusion
some final comments are made. Appendix contains some useful
technical details on gamma-matrices and Lorentz spinor harmonics.

\section{Green-Schwarz type fermionic equations of the M5 brane}
We start with the M5 brane fermionic equation in the Green--Schwarz
form
\begin{equation}\label{5.2.18}
{1\over 2}m^{ba}E_a^{\underline\beta} \left[E_b^{\underline
a}\Gamma_{\underline a}(1-\bar\Gamma)\right]
_{\underline\beta\underline\alpha} =0\,.
\end{equation}
In eq. (\ref{5.2.18})
\begin{equation}\label{pb}
E_a^{\underline\beta}=e_a^m(\xi)\,\partial_m\,Z^{\underline
M}(\xi)\,E_{\underline M}^{\underline\beta}(Z)\,,\quad
E_a^{\underline a}=e_a^m(\xi)\,\partial_m\, Z^{\underline
M}(\xi)\,E_{\underline M}^{\underline a}(Z)
\end{equation}
are the pullbacks on the M5 brane worldvolume, parametrized by the
coordinates $\xi^m$ $(m=0,1,\cdots,5)$, of the $D=11$ supergravity
supervielbeins
\begin{equation}\label{sv}
E^{\underline A}(Z)=dZ^{\underline M}\,E^{\underline
A}_{\underline M}=(E^{\underline\alpha}\,, E^{\underline a}),
\end{equation}
where $Z^{\underline M}=(x^{\underline m},\,\theta^{\underline\mu}$)
(${\underline m}=0,1,\cdots,10;~{\underline \mu}=1,\cdots,32)$ are
local coordinates of curved $D=11$ superspace \footnote{In our
notation the underlined indices correspond to $D=11$ target
superspace and not underlined ones correspond to the M5 brane
worldvolume. The indices from the beginning of the Latin and Greek
alphabet are vector and spinor tangent (super)space indices, while
the indices from the middle of the Latin and Greek alphabet are that
of local curved coordinates.}.

\noindent
 $e^m_a(\xi)$ is the inverse vielbein on the M5 brane
worldvolume associated with the induced worldvolume metric
$g_{mn}=\partial_m\,Z^{\underline M}\,E_{\underline M}^{\underline
a}\,\partial_n\,Z^{\underline N}\,E_{\underline N}^{\underline
b}\,\eta_{\underline {ab}}$. As so, $E^{~\underline a}_a$ satisfy
the orthogonality condition
\begin{equation}\label{oc}
E^{~\underline a}_a\,E^{~\underline
b}_b\,\eta_{\underline{ab}}=\eta_{ab}\,,
\end{equation}
where $\eta_{\underline{ab}}$ and $\eta_{ab}$ are respectively
$D=11$ and $d=6$ Minkowski metric.

Finally, the matrix
$m^{ab}=m^{ba}=\eta^{ab}-2\,h^{a}{}_{cd}h^{bcd}$ describes the
interaction of the fermionic (and bosonic) worldvolume fields with
the self--dual world--volume tensor
$h_{abc}={1\over{3!}}\epsilon_{abcdef}\,h^{def}$. The tensor
$h_{abc}$ is related via the nonlinear equation
\begin{equation}\label{h}
4(m^{-1})_a^{~d}\,h_{dbc}=H_{abc}\,
\end{equation}
to the field strength $H^{(3)}={1\over
{3!}}\,e^c\,e^b\,e^a\,H_{abc}=d\,b^{(2)}(\xi)-A^{(3)}$ of the M5
brane chiral 2--form gauge field $b^{(2)}(\xi)$ extended with the
pull back of the 3--form gauge potential $A^{(3)}$ of
$D=11$ supergravity whose field strength
\begin{equation}\label{fda}
F^{(4)}=dA^{(3)}= {i\over 2}\,E^{\underline a}E^{\underline b}\bar
E\Gamma_{\underline{ba}}E+ {1\over{4!}}E^{\underline
a_4}\cdots\,E^{\underline a_1}F_{\underline a_1\cdots \underline
a_4}\,
\end{equation}
generates the background fluxes. $\Gamma^{\underline
a}_{\underline\alpha\underline\beta}$ are $D=11$ gamma--matrices
in the Majorana representation and $\bar \Gamma$ is their
antisymmetrized product
\begin{equation}\label{5.2.19}
\bar\Gamma={1\over {6!}}\epsilon^{a_1\cdots \, a_6}
\Gamma_{a_1\cdots\,a_6}+{1\over 3}h^{abc}\Gamma_{abc}, \quad
\Gamma_a=E_a^{\underline a}\Gamma_{\underline a},
\end{equation}
such that $\bar\Gamma^2=1$.

 Equation (\ref{5.2.18}) is invariant under the
$\kappa$--symmetry transformations
\begin{equation}\label{4.49}
\delta_\kappa\, Z^{\underline M}\,E_{\underline M}^{\underline
\alpha}= {1\over 2}\,(1+\bar\Gamma)^{\underline
\alpha}_{~\underline\beta}\, \kappa^{\underline\beta}\,, \quad
\delta_\kappa Z^{\underline M}E_{\underline M}^{\underline a}=0\,.
\end{equation}
 Kappa--symmetry allows one to eliminate half of
the M5 brane fermionic degrees of freedom. To see this one may
notice that the right hand side of (\ref{5.2.18}) is annihilated by
the
$\kappa$--symmetry projector  $ {1\over
2}(1+\bar\Gamma)_{~~\underline\beta}^{\underline\alpha}$.

To extract the explicit dependence on fluxes in the Dirac equation
on the M5 brane in the linear approximation for the fermions we have
to evaluate the ingredients in this equation in the corresponding
approximation. We look at
\begin{equation}
E_m^{\underline \beta}=\partial_mZ^{\underline M}E_{\underline M}
^{\underline \beta}= \partial_mZ^{\underline \mu}E_{\underline
\mu} ^{\underline \beta}+ \partial_mZ^{\underline m}E_{\underline
m} ^{\underline \beta}\,. \label{vielbein}
\end{equation}
 We now define
\begin{equation}\label{tha}
\theta^{\underline \beta}\equiv Z^{\underline \mu}E_{\underline \mu}
^{\underline \beta}
\end{equation}
and rewrite eq. (\ref{vielbein}) in the following form
\begin{equation}
E_m^{\underline \beta}=\partial_m \theta ^{\underline \beta}-
Z^{\underline \mu} \partial_m E_{\underline \mu} ^{\underline
\beta}+ \partial_mZ^{\underline m}E_{\underline m} ^{\underline
\beta} \label{vielbein1}
\end{equation}
The first term is already at the linear level in $\theta$ and
provides the free Dirac equation in the flat $D=11$ background.
Without coupling to the bosonic fields and to the tensor field on
the brane we would find, for $m^{ba}=\delta^{ba}$,  $e_b^m=
\delta_b^m$ and $E_n^{\underline a}= \delta_n^a$,
\begin{equation}
E_{ a}^{\underline\beta}[ \Gamma^{ a}(1-\bar\Gamma)]_{\underline\beta\underline\alpha}
= \partial_{ a}\theta^{\underline\beta}[ \Gamma^{ a}(1-\bar\Gamma)]_{\underline\beta\underline\alpha}
=0,
\end{equation}
If we introduce the chiral spinor $\theta_-=
{1\over2}(1-\bar\Gamma)\theta$, which is manifestly invariant
under $\kappa$--symmetry transformations (\ref{4.49}), we find a
simple Dirac equation
\begin{equation}
  \Gamma^{ a}\partial_{ a}\theta_-=0
\end{equation}
This equation in turn easily transfers into a free Dirac equation
for a chiral spinor on the M5 brane. However, here we would like
to take care of corrections due to background geometry and in
particular bulk fluxes. We therefore should evaluate the remaining
terms in eq. (\ref{vielbein1}).

In the Wess--Zumino gauge the term $\partial_m E_{\underline \mu}
^{\underline \beta}$, that has to be evaluated at the zero order
in $\theta$ since it is multiplied by $\theta^{\underline \mu}$,
vanishes. The third term in eq. (\ref{vielbein1}) provides us with
the information we are looking for. The vector--spinor vielbein
$E_{\underline m} ^{\underline \beta}$ in $(11|32)$ superspace
starts with gravitino $\psi_{\underline m}^{\underline \beta}(x)$,
which we shall not take into account below restricting ourselves to
pure bosonic $D=11$ supergravity backgrounds. The term linear in
$\theta$ was calculated long time ago in \cite{Cremmer:1980ru}: this
term is proportional to the rhs of the gravitino supersymmetry
transformation excluding the term with the space--time derivative
acting on the supersymmetry parameter\footnote{The $\theta^2$ term
was given in \cite{deWit:1998tk}, the $\theta^3$ term was derived in
\cite{Grisaru:2000ij}
 and the expression for
$E^{\underline\beta}_{\underline m}$ up to the 5th order in
$\theta$ was calculated in \cite{Tsimpis:2004gq} using a compact
expression for the Wess--Zumino gauge, analogous to the one
proposed for $D=4$ supergravity in \cite{Bandos:2002bx}.}. In
notation of \cite{Sorokin:1999jx}
\begin{equation}\label{wzg}
  E_{\underline m}^{\underline \beta}=
 \psi_{\underline m}^{\underline \beta}
 +( \Omega_{\underline m  \underline \alpha }^{\underline \beta}+
 T_{\underline m  \underline \alpha}^{\underline \beta})\theta^{\underline \alpha}+...
\end{equation}
where the term linear in $\theta$ includes a $Spin(1,10)$
connection
\begin{equation}
\Omega_{\underline m  \underline \alpha }^{\underline \beta}=
({1\over 4} \Omega_{\underline m}^{  \underline a \underline
b}\,\Gamma_{\underline a \underline b })_{~\underline \alpha
}^{\underline \beta}
\end{equation}
and the flux--dependent superspace torsion term
\begin{equation}
T_{\underline m  \underline \alpha }^{\underline \beta}=
(T_{\underline m}^{~\underline{ a  b c d}}F_{\underline a \underline
b \underline c \underline d})^{\underline \beta}{}_{\underline
\alpha }\,.
\end{equation}
Here
\begin{equation}\label{T}
T^{\underline a \underline b \underline c \underline d \underline
e} \equiv {1\over 288} (\Gamma^{\underline a \underline b
\underline c \underline d \underline e} - 8 \eta^{\underline
a[\underline b}\Gamma^{ \underline c \underline d \underline e]}
)\,.
\end{equation}
Thus we get
\begin{equation}\label{Dirac}
m^{ba}\,e_b^m\,e^n_a \,\left (\partial_m\, \theta ^{\underline
\beta}+( \Omega_{ m  \underline \alpha }^{\underline \beta}+
 T_{ m  \underline \alpha}^{\underline \beta})\,\theta^{\underline \alpha}\right)
\left[E_n^{\underline a}\,\Gamma_{\underline
a}\,(1-\bar\Gamma)\right] _{\underline\beta\underline\gamma} =0,
\end{equation}

In the absence of 11D fluxes $F_{\underline a \underline b
\underline c \underline d}$ and of the M5 brane chiral gauge field,
and in the approximation in which the fermionic equation is linear
in
$\theta$, it is simply the Dirac equation with a metric compatible
spin connection  and an $SO(5)$ gauge group connection encoded in
the covariant derivative
$\nabla_a$
\begin{equation}\label{Diracfree}
 \Gamma^{a} \nabla_a \theta_- =0\,.
\end{equation}
When fluxes are present we find that
\begin{equation}\label{Diracflux}
m^{ab}\,\Gamma_{b}\,( \nabla_{a} + T_{a}{}^{\underline a \underline
b \underline c \underline d}\,F_{\underline a \underline b
\underline c \underline d}) \theta_- =0\,.
\end{equation}
Equation (\ref{Diracflux}) is of a Green--Schwarz type in the
sense that the fermionic field and the fluxes carry the target
superspace vector and spinor indices. To reduce it to an equation
which describes the dynamics of the M5 brane fermionic modes in
the effective $6d$ worldvolume field theory one can, for example,
impose on the worldvolume scalars the physical (static) gauge,
which fixes worldvolume reparametrization invariance, and
eliminate half of the fermionic modes by gauge fixing
$\kappa$--symmetry. Such a gauge fixing breaks $D=11$ local
$Spin(1,10)$ symmetry down to its subgroup $Spin(1,5)\times
Spin(5)$, where $Spin(1,5)$ is the $6d $ worldvolume 'Lorentz'
symmetry and $Spin(5)\sim USp(4)\sim SO(5)$ is the internal
R--symmetry of the effective chiral (2,0) $d=6$ supersymmetric
worldvolume field theory. This method was used in \cite{Claus:1997cq}
where the free action for the tensor multiplets on the worldvolume of M5 brane was derived.

Alternatively, one can get the same worldvolume fermion equation
in a simpler way, without breaking $D=11$ Lorentz invariance, by
singling out the irreducible $\kappa$--invariant part of
$\theta^{\underline\alpha}$ using the method of Lorentz harmonics
which is part of the superembedding approach (see
\cite{BPSTV,Sorokin:1999jx}) for a review and references). In the
next section we shall use the latter method to derive the purely
worldvolume fermionic equation with fluxes.

\section{Dirac equation on M5 surface}
Here we start with another form of the fermionic equation
\begin{equation}\label{fe}
\tilde \gamma_b^{\alpha\beta}\,m^{ba}\,E_a^{\underline\beta}\,
v_{\underline\beta,\beta q}=0\,,
\end{equation}
which is related to eq. (\ref{5.2.18}) by a certain
transformation \cite{Howe:1997fb,Sorokin:1999jx}.

In eq. (\ref{fe}) $\tilde \gamma_b^{\alpha\beta}$ and
$\gamma^{a}_{\alpha\beta}$ are antisymmetric $d=6$ $Spin(1,5)$
${\gamma}$--matrices having the properties described in eqs.
(\ref{5.2.03})--(\ref{5.2.003}) of the Appendix, and
$v_{\underline\beta,\beta q}(\xi)$ are half of the components of
the $Spin(1,10)$ matrix (called Lorentz spinor harmonics)
\begin{equation}\label{5.2.04}
v_{\underline\alpha}^{~\underline\beta}=
(v_{\underline\alpha}^{~\alpha p},v_{\underline\alpha,\beta
q})\,,\quad C^{\underline\alpha\underline\gamma}\,
v_{\underline\alpha}^{~\underline\beta}\,v_{\underline\gamma}^{~\underline\delta}
=C^{\underline\beta\underline\delta} =\left(
\begin{array}{cc}
0  &  \delta^\alpha_\beta \delta_q^{p}\\
-\delta_\gamma^\delta\delta^r_{s}   &  0
\end{array}
\right)\,.
\end{equation}
In (\ref{5.2.04}) the $Spin(1,10)$ index ${}^{\underline\beta}$ is
split into the two pairs $^{\alpha p}$ and ${}_{\beta q}$ of indices
of $Spin(1,5)\times Spin(5)$ which is the symmetry of the M5 brane
worldvolume theory. The corresponding realization of the
$D=11$ $\Gamma$--matrices is given in the Appendix, eqs.
(\ref{5.2.1})--(\ref{5.2.C}). Note that the upper and lower
$Spin(1,5)$ indices $^{\alpha}$ and ${}_{\beta}$ correspond to
inequivalent chiral spinor representations of $Spin(1,5)$, and
there is no a $6d$ charge conjugation matrix which would raise and
lower these indices.

The Lorentz harmonics (\ref{5.2.04}) are auxiliary worldvolume
fields. They are related to the pullback $E^{\underline
A}_a(Z(\xi))$ of the $D=11$ supervielbein (\ref{sv}) by the
equations (\ref{4.36})--(\ref{lvh}) of the Appendix. In
particular,
\begin{equation}\label{0}
E^{~\underline\alpha}_a\,v_{\underline \alpha}^{~\alpha p}=0\,.
\end{equation}

We can use the spinor harmonics (\ref{5.2.04}) to convert the
target space spinor field (\ref{tha}) into a pair of chiral and
anti--chiral worldvolume spinors
\begin{equation}\label{cs}
\theta^{\underline \beta}v_{\underline\beta\beta q}\equiv
\theta_{\beta q}\,, \qquad \theta^{\underline
\beta}v_{\underline\beta}^{~\alpha p}\equiv \theta^{\alpha p}\,.
\end{equation}
Because of $\kappa$--invariance (which is reflected in the
orthogonality condition (\ref{0})) the anti--chiral spinor field
$\theta^{\alpha p}$ does not appear in the fermionic equation
(\ref{fe}). In other words, one can use local $\kappa$--symmetry
transformations to put $\theta^{\alpha p}$ to zero.

Then (\ref{fe}) takes the form similar to (\ref{Diracflux}) but
with $d=6$ worldvolume  matrix $\tilde\gamma_b$ instead of pulled
back $D=11$ matrix $\Gamma_b$
\begin{equation}\label{fe1}
[\tilde\gamma_b\,m^{ba}\,(\nabla_{a} + T_{a}{}^{\underline a
\underline b \underline c \underline d}F_{\underline a \underline
b \underline c \underline d})\theta]^{\alpha}_ q=0\,.
\end{equation}
Here the covariant derivative  is the derivative with account of
metric compatible spin connection for $Spin(1,5)\times Spin(5)$
structure group (see \cite{Sorokin:1999jx} for details).

 Using the relations (\ref{5.2.1})--(\ref{lvh}) of the Appendix we can rewrite
(\ref{fe1}) in a purely worldvolume form
\begin{equation}\label{wfe}
{\tilde \gamma}^{\alpha\beta}_b\,m^{ba}\nabla_a\, \theta_{\beta
q}+{1\over{24}}\,\left[\left(\gamma^{ijk}\,
\tilde\gamma^{b}\,(2\delta^a_b-m^{~a}_b)\,F_{aijk}+\gamma^i\,
\tilde\gamma^{bcd}\,(
2\delta^a_b-3m^{~a}_b)F_{acdi}\right)\,\theta\right]^{\alpha}_q
=0\,,
\end{equation}
where the indices $i,j,k=1,2,3,4,5$ correspond to the target space
directions transversal to the M5 brane.

If we ignore the 3-form $h$ contribution, we find that in our
approximation eq. (\ref{wfe}) reduces to
\begin{equation}\label{wfe-h}
{\tilde \gamma}^{a\,\alpha\beta} \,\nabla_a\, \theta_{\beta
q}+{1\over{24}}\,\left[\left(\,\gamma^{ijk}\,
\tilde\gamma^{a}\,F_{aijk}-\,\gamma^i\,
\tilde\gamma^{abc}\,F_{abci}\right)\,\theta\right]^{\alpha}_q
=0\,.
\end{equation}

\section{ Dirac action and examples of flux transforming under $U(1)$ symmetry}

The fermion Lagrangian which produces the equations (\ref{wfe}) is
\begin{equation}\label{m5l}
L^{M5}_f={1\over 2}\,\theta\,\left[{\tilde
\gamma}_b\,m^{ba}\nabla_a\, +{1\over{24}}\,\left(\gamma^{ijk}\,
\tilde\gamma^{d}\,(2\delta^a_d-m^{~a}_d)\,F_{aijk}+\gamma^i\,
\tilde\gamma^{bcd}\,(2\delta^a_d-3m^{~a}_d)F_{abci}\right)\right]\,\theta\,.
\end{equation}

 Let us note that via $m^{ab}$, which contains the {\sl self--dual} field
$h_{abc}$ defined in (\ref{h}), the M5 brane fermions couple {\sl
directly}, though non--minimally, to the pull back of the 3--form
{\sl flux potential} $A_{abc}$. In the presence of the worldvolume
flux $h_{abc}$ both the self--dual and anti--self--dual
worldvolume parts of the flux $F_{abci}$ appear in the fermion
Lagrangian, while if we neglect the contribution of $h_{abc}$ (so
that $m^{~a}_{b}=\delta^{~a}_{b}$), the flux $F_{abci}$ should be
{\sl anti--self--dual} on the M5 worldvolume, since
$\tilde\gamma^{abc}$ is {\sl self--dual} (see eq. (\ref{5.2.0003})
of the Appendix).

The action in presence of fluxes is invariant under the $SO(5)$
structure group transformations  with the flux $F_{aijk}$
transforming as a 3d rank antisymmetric tensor and the $F_{abci}$
transforming as an $SO(5)$ vector.

We may split the $SO(5)$ index into those of $SO(3)$ and $SO(2)$,
namely $i= \mu,\nu, \lambda , I, J$. This corresponds to splitting 5
directions normal to the brane into ${\bf R}^3$ for the 3 directions
$i= \mu,\nu, \lambda$ and the remaining two directions $i= I, J$
will correspond to $SO(2)\sim U(1)$. We will be interested in the
situation that $F_{aijk}$ has only $F_{a\mu\nu\lambda}$ components
and $F_{abci}$ has only $F_{abcI}$ components. This would correspond
to the   4-fold compactification of M-theory with 8 coordinates
$a=1,..., 6, I=1,2$ to a 3-dimensional space with 3 coordinates
$\mu,\nu, \lambda$.

The action is invariant under $SO(3) \times SO(2)$ symmetry with the
flux $F_{abcI}$ transforming as a vector under $SO(2)$
symmetry.

Now we will look at an example which is more specific in the
context of M theory, orientifolds and G-flux vacua
\cite{Dasgupta:1999ss}, \cite{Gorlich:2004qm}. In M-theory one
starts with $X=K3_1\times K3_2$ four-fold in the presence of the
flux $F_4$. In type IIB this is an orientifold $K3\times {T^2\over
Z_2}$. This $K3$ is $K3_1$ in M-theory. The second $K3$ which is
called $K3_2$ is elliptically fibered.  Thus we have an M-theory
on a Calabi-Yau four-fold ${\bf R}^3 \times X$. We will take an
Euclidean signature both in the target as well as on M5. In this
case the complex divisor $D$ on which the five-brane is wrapped is
a six-dimensional cycle in $X$. The 4-flux related to a 3-form
flux in type IIB theory $G_3= F_3-\phi H_3$ (where $F_3$ and $H_3$
are  respectively RR and NS 3-forms and $\phi$ is a complex
axion--dilaton of IIB theory) can be chosen as follows
\begin{equation}
F_4= -{1\over \phi-\bar \phi}\, G_3 \wedge d\bar z_2+ {1\over
\phi-\bar \phi} \bar G_3\, \wedge d z_2 \label{4form}\,,
\end{equation}
where $dz_2$ and $d\bar z_2 $ are holomorphic and anti-holomorphic
differentials along the elliptically fibered torus in $K3_2$.

The examples of supersymmetric background fluxes studied in
\cite{Dasgupta:1999ss} and \cite{Gorlich:2004qm} require that $F_4$
has two legs along $K3_1$ and two legs along $K3_2$. This means that
$H_3$ and $F_3$ have 2 legs in $K3_1$ and one leg in $K3_2$,
in direction with coordinates $z_1, \bar z_1$. In our
split of the 11-dimensional Euclidean space into 6+5, the six
coordinates of the M5 have to be in $D$. The 5 directions normal to
the brane include
${\bf R}^3$ for  the 3 directions. The remaining 2 directions,
normal to $D$ are related to the $SO(2)\sim U(1)$ symmetry, which is
a rotation  in $z_1,  \bar z_1$ plane.  One of the examples
studied in \cite{Dasgupta:1999ss} and \cite{Gorlich:2004qm} is
\begin{equation}
F_4 = C  \Omega \wedge d\bar z_1 \wedge d\bar z_2+ \bar C \bar
\Omega \wedge dz_1\wedge d z_2  \label{example}\,,
\end{equation}
where $C$ is a complex constant and where $\Omega$ ($\bar \Omega$)
is a holomorphic  (anti-holomorphic) two-form on $K3_1$. The flux
has an $F_{z_1}$ as well as an  $F_{\bar z_1}$ component which
 under the
$U(1)$ transformation with a parameter $\varphi$ acquire the phase
$e^{i\varphi} F_{z_1}$ and $e^{-i\varphi} F_{\bar z_1}$, respectively.  If we
would consider the fixed vacuum expectation value of the flux we
would see that it violates the $U(1)$ symmetry. However, in the
context  in which the background flux transforms under $U(1)$ we
have the following situation. The flux transforms under $U(1)$ if
$C$ transforms as $e^{-i\varphi} C$ and $\bar C$ as $e^{i\varphi}
\bar C$.
 This leaves us with the Dirac action on M5
in presence of the background fluxes which is invariant under
$U(1)$.

\section{The Dirac equation and action on the M2 brane in the presence of $D=11$ fluxes}
For completeness, here we present the Green--Schwarz and purely
worldvolume form of the Dirac operator on an M2 brane coupled to a
$D=11$ supergravity gauge field flux $F_{\underline{abcd}}$. Its
derivation can be carried out in the same way as for the M5 brane.
In the Green--Schwarz form the M2 brane fermionic equation
\cite{Bergshoeff:1987cm} is
\begin{equation}\label{gsm2}
\eta^{ab}\,E^{\underline\beta}_a\,\left[E^{\underline
a}_b\Gamma_{\underline
a}\,(1-\bar\Gamma)\right]_{\underline{\beta\alpha}}=0\,,
\end{equation}
where
$\bar\Gamma={1\over{3!}}\,\epsilon^{abc}\,\Gamma_{abc},~\bar\Gamma^2=1$,
$\Gamma_a=E^{\underline\beta}_a\,\Gamma_{\underline a}$ and $a=0,1,2$
are the worldvolume tangent space indices.

Using eqs. (\ref{wzg})--(\ref{T}) we find that in the linear
approximation in $\theta$ eq. (\ref{gsm2}) reduces to
\begin{equation}\label{Diracfluxm2}
\Gamma^{a}\,( \nabla_{a} +   T_{a}{}^{\underline  a \underline b
\underline c \underline d}\,F_{\underline a \underline b \underline
c \underline d})\, \theta_- =0\,,
\end{equation}
where $\theta_-={1\over2}(1-\bar\Gamma)\theta$, being
`kappa--projected', has 16 independent components. This form of
equation can also be extracted from the more general answer in
\cite{deWit:1998tk}  or from the  M2 brane quadratic
action of \cite{Marolf:2003vf,Marolf:2003ye}.

To get the purely worldvolume form of the M2 brane Dirac operator
with fluxes, one  starts from the
$M2$ brane fermionic equation in the superembedding formulation
\cite{BPSTV,Sorokin:1999jx}
\begin{equation}\label{m2f}
\gamma^{a\alpha\beta}\,E_a^{\underline\beta}\,
v_{\underline\beta,\,\beta q'}=0\,,
\end{equation}
where now $\gamma^{a}_{\alpha\beta}=\gamma^{a}_{\beta\alpha}$ are
$d=3$ $M2$ worldvolume symmetric gamma--matrices whose spinor
indices $\alpha,\beta=1,2$ are raised and lowered by the
antisymmetric unit matrices
$\epsilon^{\alpha\beta}=\epsilon_{\alpha\beta}$, $q'=1,\cdots,\,8$
is the index of a spinor representation of the $SO(8)$ group of
transformations of $d=8$ target space directions transversal to
the $M2$ brane and $v_{\underline\beta,\beta q'}$ are half of the
components of the $Spin(1,10)$ spinor Lorentz harmonics. Then,
using eqs. (\ref{wzg})--(\ref{T}) and expressions of Section 5.1
of \cite{Sorokin:1999jx}, one can reduce eq. (\ref{m2f}) to a form
analogous to that for the M5 brane
\begin{equation}\label{m2ff}
\gamma^{a\,\alpha\beta} \,\nabla_a\, \theta_{\beta
q'}+{1\over{96}}\,\left[\left(\gamma^{ijkl}\,
F_{ijkl}-6\,\epsilon^{abc}\,\gamma_c\,\gamma^{ij}
\,F_{abij}\right)\,\theta\right]^{\alpha}_{q'} =0\,.
\end{equation}
The corresponding $M2$ brane worldvolume Lagrangian is
\begin{equation}\label{m2l}
L^{M2}_f={1\over 2}\, \theta\,\left[\gamma^{a} \,\nabla_a\,
+{1\over{96}}\,\left(\,\gamma^{ijkl}\, F_{ijkl}-6
   \,\epsilon^{abc}\,\gamma_c\,\gamma^{ij}
\,F_{abij}\right)\right]\,\theta\,.
\end{equation}
In eqs. (\ref{m2ff}) and (\ref{m2l})
$\gamma^{ijkl}_{q'p'}=\gamma^{ijkl}_{p'q'}$ and
$\gamma^{ij}_{q'p'}=-\gamma^{ij}_{p'q'}$ are antisymmetric
products of $SO(8)$ gamma matrices $\tilde\gamma^i_{q'p}$ and
$\gamma^i_{pr'}$ with the indices $i,j,k,l=1,\cdots\,,8$ labeling
the vector representation of $SO(8)$ and the index
$p=1,\cdots\,,8$ corresponding to the second  spinor
representation of $SO(8)$ different from that labeled by $q'$
 (see e.g. \cite{BPSTV,Sorokin:1999jx} for details). Note that there is no difference between
 upper and lower $SO(8)$ indices since they all are raised and lowered by the unit symmetric matrices
 $\delta^{ij}$, $\delta^{pq}$ and $\delta^{p'q'}$.  The gamma matrices
 $\gamma^i_{pr'}=\tilde\gamma^i_{r'p}$, such that
  $\gamma^i_{pr'}\tilde\gamma^j_{r'q}+\gamma^j_{pr'}\tilde\gamma^i_{r'q}=\delta^{ij}\delta_{pq}$,
imply well known triality of the three inequivalent 8--dimensional
fundamental representations of $SO(8)$.

 It is interesting to note that in the presence of
fluxes there is a kind of anomalous magnetic moment coupling of
the worldvolume fermions to the field strengths of the fluxes. In
the most straightforward way this anomalous magnetic moment
coupling is seen in the Dirac equation for the M2 brane
(\ref{m2ff}). In the last term of this equation $F_{abij}$ can be
regarded as a 2--form field strength on the 3d worldvolume with
$i,j$ being the indices of an internal local symmetry group. When
$i,j=1,2$ take only the values of $SO(2)$, we see that this is
nothing but the anomalous magnetic moment coupling of 3d fermions
to an electromagnetic field strength in a 3d field theory. A
similar interpretation may also be given to flux terms in the
Dirac operator on the M5 brane.  For instance, the first flux
term in (\ref{m5l}) can be rewritten as $\gamma^{ijk}\,
\tilde\gamma^{a}(2\delta^d_a-m^{~d}_a)\,F_{dijk}=\tilde\gamma^{a}\,F_{a}^{ij}\,\gamma_{ij}$
(where $F_a^{ij}={1\over
{3!}}\,\varepsilon^{ij\,i_1j_1k_1}\,(2\delta^a_d-m^{~a}_d)\,F_{d\,i_1j_1k_1}$
and $\gamma_{ij}={1\over
{3!}}\,\varepsilon_{ij\,i_1j_1k_1}\,\gamma^{i_1j_1k_1}$). One can
notice that the `anomalous magnetic moment' term
$F_{a}^{ij}\,\gamma_{ij}$ has the form similar to that of the
$SO(5)$ connection $\Omega^{ij}_a\,\gamma_{ij}$ which enters the
covariant derivative of the M5 brane Dirac operator
(\ref{m5l}).

\section{Conclusion}

Thus we have derived here the explicit dependence on fluxes in the
fermionic action on the M5 brane in a generic background.  We
have shown that there are two types of fluxes which enter the Dirac
action. One of them transforms as an antisymmetric rank 3 tensor and
another one as a vector of the R--symmetry group $SO(5)$ of the
M5-brane, and in particular of its SO(2) subgroup.
  This poses a question: what happens in
general with the condition $\chi_D=1$ derived in
\cite{Witten:1996bn} and studied more recently in
\cite{Robbins:2004hx} and  \cite{Gorlich:2004qm}. In examples of
compactification shown in \cite{Gorlich:2004qm} the answer was that
$\chi_D\geq 1$ might provide the non-vanishing superpotential. It is
not yet known how to  generalize Witten's analysis for the Dirac
operator in eq. (\ref{wfe}).  This equation is a generalization
of a simple Dirac equation used in \cite{Witten:1996bn}  under
the condition that there is no background flux, $F_4=0$ and there is
no chiral 2-form on M5, $b_2=0$. When the background fluxes and
world-volume 2-forms are present, in general, the analysis of the
instanton corrections has to be redone. The immediate reason for
this is the fact that background fluxes are necessarily required for
stabilization of the dilaton-axion and complex structure moduli
\cite{GKP}.

The importance of this topic has to do with the fact that the only
known way at present in which string theory and higher-dimensional
supergravities may, possibly, address the current cosmological
observations, require stabilization of moduli.  The most difficult
part, stabilization of K\"ahler moduli, is based on non-perturbative
instanton corrections to the superpotential discussed here.
Clarification of the restrictions on compactification manifolds
which might provide such non-perturbative superpotentials  might
lead to a significant  progress in string cosmology.

\vskip0.5cm

\leftline{\bf Acknowledgements} We are grateful to F. Denef and B.
Florea who suggested to look for the Dirac operator with fluxes on  M5
 in the context of
instanton corrections, and to E. Bergshoeff, I. Bandos, J. Gomis, S. Gukov,
 S. Kachru, G. Moore, A. Kashani-Poor, and
 S. Sethi, A. Tomasiello, S. Trivedi and A. Van Proeyen for
clarifying discussions. The work of R. K. was supported by NSF
grant 0244728. The work of D. S. was supported by the  EU
MRTN-CT-2004-005104 grant `Forces Universe', and by the MIUR 
contract no. 2003023852.

\section*{Appendix}
In our notation the underlined indices correspond to $D=11$ target
superspace and not underlined ones correspond to the M5 brane
worldvolume. The indices from the beginning of the Latin and Greek
alphabet are vector and spinor tangent (super)space indices, while
the indices from the middle of the Latin and Greek alphabet are that
of local curved coordinates. The letters $i,j,k$ and
$p,q,r,s$ stand, respectively for vector and spinor $Spin(5)$
indices.

We use the form of the $D=11$ $\Gamma$--matrices and of the charge
conjugation matrices
$C_{\underline\alpha\underline\beta}=C^{\underline\alpha\underline\beta}$
which reflects the embedding of the M5 brane $6d$ worldvolume into
$N=1$, $D=11$ superspace
\begin{equation}\label{5.2.1}
\Gamma^a_{\underline\alpha\underline\beta}=\left(
\begin{array}{cc}
\gamma^a_{\alpha\beta}C_{pq}  &  0\\
0  &  \tilde\gamma^{a\alpha\beta}C^{pq}
\end{array}
\right)\,, \quad a=0,1,...,5, \quad \alpha,\beta=1,2,3,4,
\end{equation}
\begin{equation}\label{5.2.2}
\Gamma^i_{\underline\alpha\underline\beta}=\left(
\begin{array}{cc}
0  &  \delta^\alpha_\beta(\gamma^{i})_q^{~p}\\
-\delta_\alpha^\beta(\gamma^i)^q_{~p}   &  0
\end{array}
\right)\,, \quad i=1,...,5\,,\quad q,p=1,2,3,4\,,
\end{equation}
\begin{equation}\label{5.2.C}
C_{\underline\alpha\underline\beta}=C^{\underline\alpha\underline\beta}
=\left(
\begin{array}{cc}
0  &  \delta^\alpha_\beta \delta_q^{p}\\
-\delta_\alpha^\beta\delta^q_{p}   &  0
\end{array}
\right)\,.
\end{equation}
In (\ref{5.2.2})
$(\gamma^i)^q_{~p}=C^{qs}(\gamma^{i})_s^{~r}C_{rp}$ are
$USp(4)\sim SO(5)$ gamma--matrices and $C^{qs}=C_{qs}$ are charge
conjugation matrices. The matrices
$(\gamma^{i})_{qp}=(\gamma^{i})_q^{~r}C_{rp}$ and $C^{qs}$ are
antisymmetric.

$\tilde \gamma_b^{\alpha\beta}$ and $\gamma^{a}_{\alpha\beta}$ are
antisymmetric $d=6$ $Spin(1,5)$ ${\gamma}$--matrices having the
following properties
\begin{equation}\label{5.2.03}
\gamma^a_{\alpha\gamma}\tilde\gamma^{b\gamma\beta}+
\gamma^b_{\alpha\gamma}\tilde\gamma^{a\gamma\beta}
=2\delta^{\beta}_\alpha\eta^{ab}, \qquad {\rm
tr}(\gamma^a\tilde\gamma^b)=4\eta^{ab}, \quad
\gamma_{a\alpha\beta}\gamma^a_{\gamma\delta}
=-2\epsilon_{\alpha\beta\gamma\delta}\,,
\end{equation}
\begin{equation}\label{5.2.003}
\gamma^{abc}_{\alpha\beta}= \gamma^{abc}_{\beta\alpha}\equiv
(\gamma^{[a}\,\tilde\gamma^{b} \,\gamma^{c]})_{\alpha\beta}
 =-{1\over 6}\epsilon^{abcdef}(\gamma_{def})_{\alpha\beta}\,,
\end{equation}
\begin{equation}\label{5.2.0003}
\tilde\gamma^{abc\,\alpha\beta}=
\tilde\gamma^{abc\,\beta\alpha}\equiv
(\tilde\gamma^{[a}\,\gamma^{b}\,
\tilde\gamma^{c]})^{\alpha\beta}={1\over
6}\epsilon^{abcdef}\,\tilde\gamma_{def}^{\alpha\beta}\,,
\end{equation}

The Lorentz spinor harmonics (\ref{5.2.04})
$$
v_{\underline\alpha}^{~\underline\beta}=
(v_{\underline\alpha}^{~\alpha p},v_{\underline\alpha,\beta
q})\,,\qquad C^{\underline\alpha\underline\gamma}\,
v_{\underline\alpha}^{~\underline\beta}\,v_{\underline\gamma}^{~\underline\delta}
=C^{\underline\beta\underline\delta}\,, \qquad
v^{\underline{\alpha}}{}_{\underline{\gamma}}
=C^{\underline\alpha\underline\alpha'}\,v_{\underline\alpha'}{}^{\underline\gamma'}\,
C_{\underline\gamma'\underline\gamma}\,$$
 are auxiliary worldvolume
fields. They are related to the pullback $E^{\underline
A}_a(Z(\xi))$ of the $D=11$ supervielbein (\ref{sv}) by the
following equations
\begin{eqnarray}\label{4.36}
E^{~\underline\alpha}_a\,v_{\underline \alpha}^{~\alpha p}=0\,,
\qquad
 \Gamma^{a}_{\underline{\gamma}\underline{\delta}}\,
E^{~\underline{a}}_{{a}}+\Gamma^{i}_{\underline{\gamma}\underline{\delta}}\,u_i^{~\underline
a}\equiv\Gamma^{\underline
b}_{\underline{\gamma}\underline{\delta}}\,u_{\underline
b}^{~\underline a}
 = v^{\underline{\alpha}}{}_{\underline{\gamma}}\,
\Gamma^{{\underline a}}_{\underline{\alpha}\underline{\beta}}\,
v^{\underline{\beta}}{}_{\underline{\delta}}\,,
\end{eqnarray}
where $u_i^{~\underline a}(\xi)$  are a set of five $D=11$ Lorentz
vectors with indices ($i=1,\cdots,\,5$) belonging to the vector
representation of $Spin(5)$. $u_i^{~\underline a}(\xi)$ are
defined to be orthogonal to $E^{~\underline{a}}_{{a}}$, i.e.
$$
E^{~\underline{a}}_{{a}}\,u_i^{~\underline
b}\,\eta_{\underline{ab}}=0\,,
$$
and complement the latter to an $SO(1,10)$ matrix (also called
Lorentz vector harmonics)
\begin{equation}\label{lvh}
u_{\underline b}^{~\underline a}=(E_a^{~\underline
a},\,u_i^{~\underline a}),\qquad u_{\underline b}^{~\underline
a}\,u_{\underline d}^{~\underline
c}\,\eta_{\underline{ac}}=\eta_{\underline{bd}}\,, \qquad
u_{\underline b}^{~\underline a}\,u_{\underline d}^{~\underline
c}\,\eta^{\underline{bd}}=\eta^{\underline{ac}}\,.
\end{equation}
Using the relations (\ref{5.2.1})--(\ref{lvh}) one can show that
the fermionic equations (\ref{5.2.18}) and (\ref{fe}) are
equivalent \cite{Howe:1997fb,Sorokin:1999jx}, with the projector
matrix ${1\over 2}(1-\bar\Gamma)$ (\ref{5.2.19}) having the
following form in terms of the Lorentz spinor harmonics
(\ref{5.2.04}) and the self--dual tensor field $h_{abc}$
\begin{equation}\label{pro}
{1\over 2}(1-\bar\Gamma)_{\underline\alpha\underline\beta}=
v_{\underline\alpha}^{~\beta p}(v_{\underline\beta,\,\beta p}
+C_{pq}\,h_{abc}\,\gamma^{abc}_{\beta\gamma}\,v^{~\gamma
q}_{\underline\beta})\,.
\end{equation}

\end{document}